\documentclass[]{spie}  %>>> use for US letter paper
\usepackage{amsmath,amsfonts,amssymb}
\usepackage{graphicx}
\usepackage[colorlinks=true, allcolors=blue]{hyperref}

\title{Eliminating stray radiation inside large area imaging arrays}

\author[a]{Stephen J. C. Yates}
\author[b]{Simon Doyle}
\author[b]{Peter Barry}
\author[a,c]{Andrey M. Baryshev}
\author[d]{Juan Bueno}
\author[a]{Lorenza Ferrari}
\author[e]{Nuria Llombart}
\author[d]{Vignesh Murugesan}
\author[e]{David J. Thoen}
\author[e]{Ozan Yurduseven}
\author[d,e]{Jochem J. A. Baselmans} 
\affil[a]{SRON, Landleven 12, 9747 AD Groningen, The Netherlands}
\affil[b]{School of Physics and Astronomy, Cardiff University, Cardiff CF24 3AA, UK}
\affil[c]{Kapteyn Astronomical Institute, University of Groningen, P.O. Box 800, 9700 AV Groningen, The Netherlands}
\affil[d]{SRON, Sorbonnelaan 2, 3584 CA Utrecht, The Netherlands}
\affil[e]{Terahertz Sensing Group, Faculty of Electrical Engineering, Mathematics and Computer Science, Delft University of Technology, Mekelweg 4, 2628 CD Delft, The Netherlands}

\authorinfo{Further author information: (Send correspondence to S.Yates)\\S.Yates.: E-mail: s.yates@sron.nl\\
P. Barry now with Kavli Institute for Cosmological Physics, University of Chicago, 5640 South Ellis Avenue Chicago, IL 60637, USA}

\begin{document} 
\maketitle

\begin{abstract}
With increasing array size, it is increasingly important to control stray radiation inside the detector chips themselves. We demonstrate this effect with focal plane arrays of absorber coupled Lumped Element microwave Kinetic Inductance Detectors (LEKIDs) and lens-antenna coupled distributed quarter wavelength Microwave Kinetic Inductance Detectors (MKIDs). In these arrays the response from a point source at the pixel position is at a similar level to the stray response integrated over the entire chip area. For the antenna coupled arrays, we show that this effect can be suppressed by incorporating an on-chip stray light absorber. A similar method should be possible with the LEKID array, especially when they are lens coupled.
\end{abstract}


\section{INTRODUCTION}
\label{sec:intro}  % \label{} allows reference to this section
Large ultra-sensitive detector arrays are needed for present and future observatories for far infra-red, submillimeter wave (THz), and millimeter wave astronomy. Example applications include wide field cameras such as NIKA2~\cite{NIKA2:AA18short}, in development projects such as A-MKID~\cite{amkid} and MUSCAT~\cite{Brien:LTD18} and future requirements of large CMB experiments~\cite{CMBS4reviewshort}. Since detector substrates are transparent, and given that no detector has 100~\% optical efficiency it is clear that some radiation will be scattered and confined in the detector chip, due to its high refractive index, which will then lead to a stray light path. We demonstrate this effect with focal plane arrays of Lumped Element microwave Kinetic Inductance Detectors (LEKIDs~\cite{Doyle:JLTP08}) and lens-antenna coupled~\cite{yates:IEEETTT18short} Microwave Kinetic Inductance Detectors (MKIDs~\cite{day03}). Presented here are near field measurements of the MKID optical response versus the position on the array of a reimaged optical source. In these arrays the response from a point source at the pixel position is at a similar level to the stray response integrated over the entire chip area. For the antenna coupled arrays, we have shown~\cite{yates:IEEETTT18short} that this effect can be suppressed by a factor $>10$ by incorporating an on-chip stray light absorber. 

\section{Arrays under test}
We present here results on three arrays: one classic LEKID~\cite{Doyle:JLTP08} array which is compared to lens antenna arrays with and without an on chip stray light absorber.
\subsection{LEKID array}
\begin{figure*}[hbt]
\centering
\includegraphics[width=0.9\textwidth]{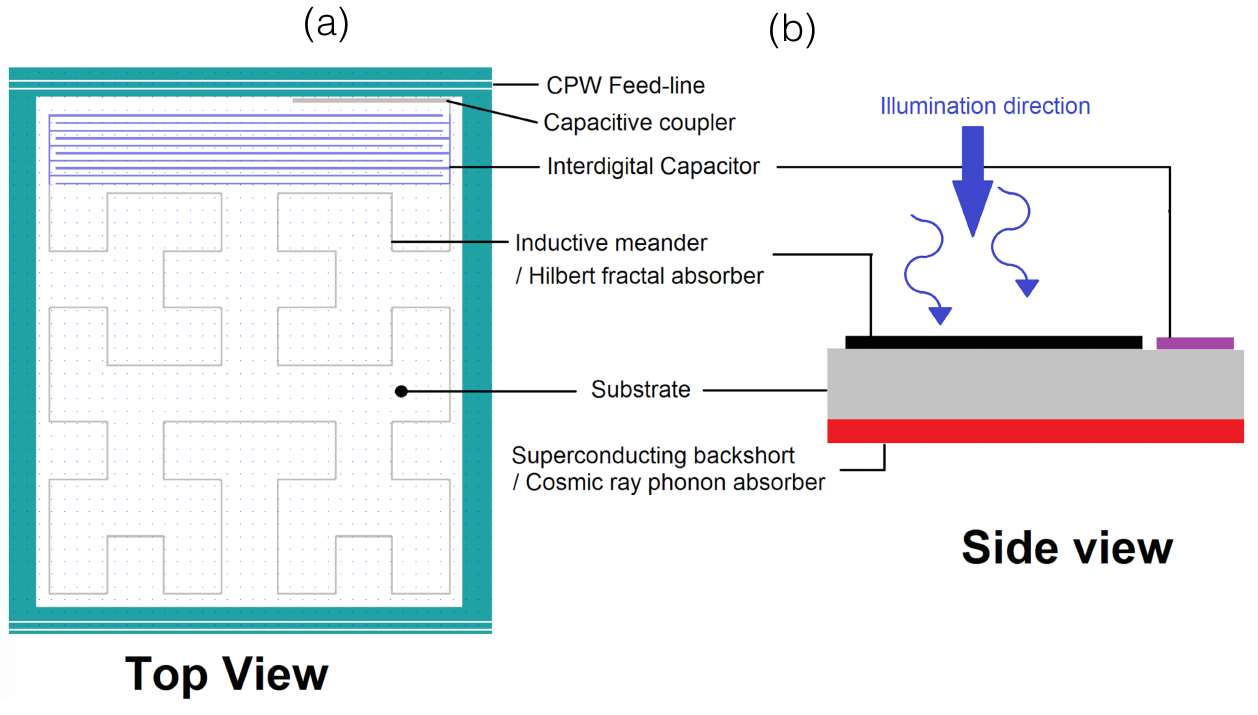}
\caption{a) Schematic of one LEKID, showing the inductor which is a dual polarization Hilbert fractal absorber, interdigital capacitor and CPW feed line. Some of the ground plane is left around the pixel, acting as a screen to reduce crosstalk. The CPW feedline is balanced with bondwires (not shown). (b) Side view schematic of the detector chip.}
\label{fig:LEKID_photo}
\end{figure*} 

The lumped element kinetic inductance detector (LEKID~\cite{Doyle:JLTP08}) array is fabricated from aluminum on silicon substrate. Each pixel is made of a lumped element interdigitated capacitor and a meander inductor, where radiation is directly absorbed in the inductor. The inductor design uses a dual polarization third order Hilbert fractal design~\cite{NIKA2:AA18short}, shown in Fig.~\ref{fig:LEKID_photo} and of size 1.28$\times$1.28~mm. The smallest feature size is 4~$\mu$~m, and the film thickness is $\sim 30$~nm. The array is front illuminated, with the backside coated in aluminum to act as a backshort. The substrate thickness is near $3\lambda/4$ at 350~GHz. A larger bandwidth can be gained with a $\lambda/4$ backshort, but for ease of handling the thicker value is taken with the consequence of lowering the average in band optical efficiency. The capacitor is varied for each pixel, varying the resonance frequency from 1.9--3.5~GHz. A CPW throughline passes by each pixel, coupling via the distance to the LEKID capacitive coupler. The throughline is balanced by placing Al wiring bonds over it, and around each pixel a continuous ground plane "screen" is present. The pixel encoding ensures that nearest neighbors in readout frequency are not physical neighbors. These strategies ensure low pixel-pixel crosstalk compared to that seen in some LEKID arrays~\cite{omid:IEEEMTT12}: crosstalk is now limited to overlapping resonators. The full pixel footprint, including meander, capacitor, screen and throughline is 1.46~mm $\times1.69$~mm. The array has 625 pixels, square packed connected to a single feedline. The pixels cover 38~mm $\times$40.6~mm of a 55~mm $\times$55~mm chip.  We should note, this array qualitatively similar to those developed for NIKA and NIKA2~\cite{NIKA2:AA18short}, except optimized for 350~GHz operation.

\subsection{Lens-antenna coupled MKIDs}
\begin{figure*}[hbt]
\centering
\includegraphics[width=1\textwidth]{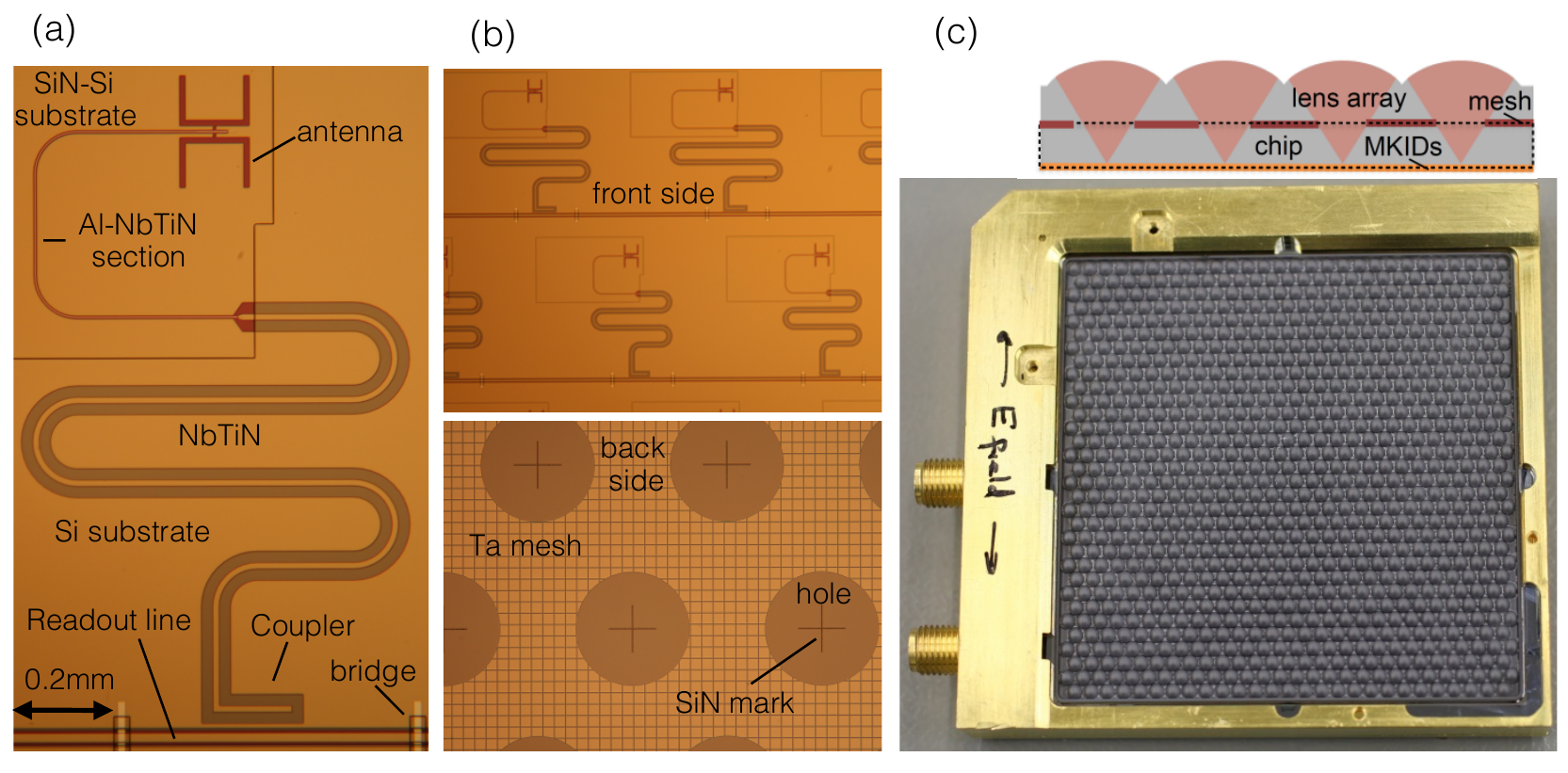}
\caption{Figure a summary modified from~\cite{yates:IEEETTT18short}, a) Optical micrograph of a single pixel of the array, artificial coloring is used to highlight the different metals.(b) Optical micrograph of the array, with the front side (top) and backside (bottom) showing the detectors on the front side and the Ta absorbing mesh on the backside, implemented on only one of the two arrays discussed in the text. (c) Assembled detector holder with lens array and SMA connector for contacting the readout circuitry. The top panel shows schematically the assembled cross section.}
\label{fig:LA_photo}
\end{figure*} 

We compare the LEKID results to previously published results~\cite{yates:IEEETTT18short} on hybrid NbTiN-Aluminum lens-antenna coupled MKIDs, which are presented in detail in previous papers~\cite{yates:IEEETTT18short,Janssen:APL13,ferrari:IEEETTT18}. To summarize we describe the detector geometery used in this work: each detector consists of a meandering coplanar waveguide (CPW) with an open end near a readout line and a shorted end at the location of the antenna. Radiation is focused by an elliptical lens onto a twinslot antenna~\cite{Jonas:IEEETMTT92,filipovic:IEEETMTT93}, which couples the radiation into the MKID CPW line. The radiation is absorbed in the first $\sim$1.5~mm section of the MKID, which is narrow with the central line of the CPW made out of 55~nm thick sputter deposited aluminum (linewidth = 1.6 $\mu$m, gapwidth = 2.2~$\mu$m) to enhance the device response and optical efficiency. The rest of the MKID and ground plane is made out of a 500~nm thick sputtered film of NbTiN, which is a wide section with a central line width = 14~$\mu$m and a central line to ground plane gapwidth = 24 $\mu$m. Here, additionally a protective SiN layer is removed prior to the NbTiN deposition. Both of these are done to reduce excess device noise due to two level systems associated with the amorphous SiN~\cite{Gao2008b}. Radiation coupled to the antenna is transferred to the narrow NbTiN-Al CPW line of the MKID and absorbed only in the aluminum central strip of the MKID: the gap frequency of NbTiN does not allow for radiation absorption below 1.1~THz whereas aluminum absorbs radiation for frequencies in excess of 90~GHz. Additionally the higher gap of NbTiN confines the optically generated quasiparticles in the Al section, further boosting optical response. The antenna is optimized for single mode single polarization radiation coupling in a 60~GHz band around 350~GHz~\cite{Ozan2016}. The antenna efficiently couples to the lens, coupling 82\% of the power of the detector beam pattern. However, the MKID itself will weakly couple to stray radiation in the substrate: due to the meander shape this will be in both polarizations. 

The two lens-antenna arrays we consider in this paper both consist of 880 pixels hexagonally packed with a pixel spacing of 2 mm, covering an area of 55.7~mm$\times56$~mm on a 62~mm$\times60.8$~mm chip. Across the array, the MKID length {\it L} is changed systematically from 6.6 to 3.5~mm, resulting in F$_0$ ranging from 4.2 to 7.8~GHz on a single readout line. We use aluminum bridges with lithographically-defined polyimide supports to balance the two grounds of the readout line. Additionally, we spatially encode the pixels such that neighboring pixels are separated sufficiently in readout frequency. Both techniques reduce MKID-MKID crosstalk~\cite{Yates:JLTP14}. Residual crosstalk is now limited by resonator overlap~\cite{Adane:JLTP16,Bisigello:SPIE16,baselmans:AA17short}, limited in our case by the NbTiN film uniformity~\cite{Thoen2016} and the MKID Q-factors under operation. Efficient radiation coupling to the MKID antennas is achieved by using a Si lens array of spherical lenses fabricated using laser ablation from a separate Si wafer. The lens array and chip are mounted together using a dedicated alignment and bonding technique where the lens array and chip are pressed together using a silicone-based press system before a semi-permanent bond is made using Locktite 406 glue. This method guarantees a glue gap below 5~$\mu$m over the entire chip area. Alignment is achieved by markers in the SiN layer on the detector chip backside that were etched in the first step of the device fabrication. The large area of the chip requires the lens array and the detector chip to be made from the same material to guarantee reliable bonding during thermal cycling of the detector assembly. The detector chip is mounted in a dedicated holder and wire bonding is used to contact the two bond pads to standard SMA co-axial connectors.

The second lens-antenna array is equipped with a stray radiation absorbing layer, on the backside of the detector chip, fabricated from a sputtered 40~nm thick Ta layer. The Ta grows in its $\beta$-phase \cite{schrey1970}, characterized by a high resistivity and low critical temperature \cite{mohazzab2000}. For our film we measure a sheet resistance $\mathrm{R_s=61~\Omega/\square\;and\;T_{c}=0.65~K}$; it is noteworthy that this observed sheet resistance gives the maximum radiation absorption for a metal layer in between two Si substrates (i.e. the lens array and the chip). The gap frequency of the Ta layer is approximately 50~GHz, such that it is superconducting at the read out frequency and resistive at 350 GHz with a resistivity very close to the normal state resistance. Using a parametric sweep, the mesh design is optimized for maximum radiation absorption upto large angles for both the TE and TM mode at 350~GHz and for maximum transmission at the MKID readout frequency of 4--8~GHz. The transparency from 4--8~GHz is needed because the mesh is only 350~$\mu$m distance from the MKIDs and therefore close enough to couple to the device. Without this the MKID will be sensitive to power absorbed in the stray light absorbing layer and would additionally have an enhanced coupling to the readout line. To efficiently couple the radiation from the lenses to the antenna, a 1.1~mm diameter hole is etched in the mesh. The mesh is a small perturbation on the beam pattern, reducing the calculated lens-antenna aperture efficiency from 0.75 to 0.74. 

\section{Array efficiency}
We define the array efficiency to be the fraction of the total power incident on the array directly absorbed by the pixels.
For a multi-moded plane absorber like the LEKID, this is to first order the filling fraction of the absorber multiplied by the optical efficiency of a single absorber over the entire frequency band and opening angles. However, the array efficiency for a single moded detector (such as one using an antenna) is a function of the single pixel optical efficiency and of the spatial sampling~\cite{Griffin:AO02}: at high sampling you can absorb all the modes falling on the array, while at low sampling you miss these modes and the associated power. For many instruments, spatial undersampling is a convenient design parameter that can be modified to balance system resources while enabling filling of the required field of view~\cite{Griffin:AO02}: with single moded detectors this gives the previously described deterioration in observing speed; however in multi-moded (LEKID) undersampling gives a deterioration in the on sky angular resolution as the physical size of the absorber is convolved in the final instrument beam. All arrays presented have a similar spatial sampling with pixel sizes of $\sim1f\#\lambda$, f\# the f/D ratio of the optics, undersampled by two times from Nyquist sampling. Additionally, the lens-antenna is single polarization sensitive, while the LEKID is dual polarization so giving a factor two in difference in efficiency, since the measurements are from an unpolarized thermal source.

For the LEKID array, the packing efficiency of the radiation absorbing inductor as 0.65. In a separate measurement, the single pixel dual polarization optical efficiency is measured as $\sim 35\%$ from the temperature dependent photon noise from a blackbody source, where this takes the inductor as the pixel size. 
This is low due to the narrow bandwidth from using a $3/4\lambda$ backshort. The total array optical efficiency including packing efficiency is therefore $\eta_{array}|_{LK}=0.22$. 

For the lens-antenna array, the beam is calculated to be coupled to the instrument focal plane with an efficiency $\eta_{loss}=0.44$, given by the spatial sampling which controls the beam truncation at the cold stop~\cite{Griffin:AO02}. Similar pixel designs have verified the beam pattern calculation, showing a high optical efficiency of the full lens-antenna beam of ($\sim74\%$)~\cite{ferrari:IEEETTT18}. The beam is a single mode of optical throughput $\lambda^2$, $ \lambda=0.86$~mm, so each beam has a transmitted throughput of $\eta_{loss} \lambda^2$. The physical individual pixel throughput is $A\Omega$ where $A$ the pixel area and $\Omega$ the solid angle. For f\#2 optics, 2~mm diameter pixel then $A\Omega=0.6$~mm.str$^2$. Therefore with a hexagonal packing efficiency $\eta_{hex}=0.92$, the lens-antenna array efficiency is the ratio of the transmitted single moded throughput to the physical throughput: 
\begin{equation}
\eta_{array}|_{l-a}=\eta_{hex}\frac{\eta_{loss}\lambda^2}{A\Omega}=0.24
\end{equation}
We see that the LEKID and lens-antenna arrays have a similar dual polarization array efficiency. These numbers do not include stray light coupling, which will be present in the arrays without a straylight absorber.

\section{System beam map}
\begin{figure}
\centering
\includegraphics*[width=10cm]{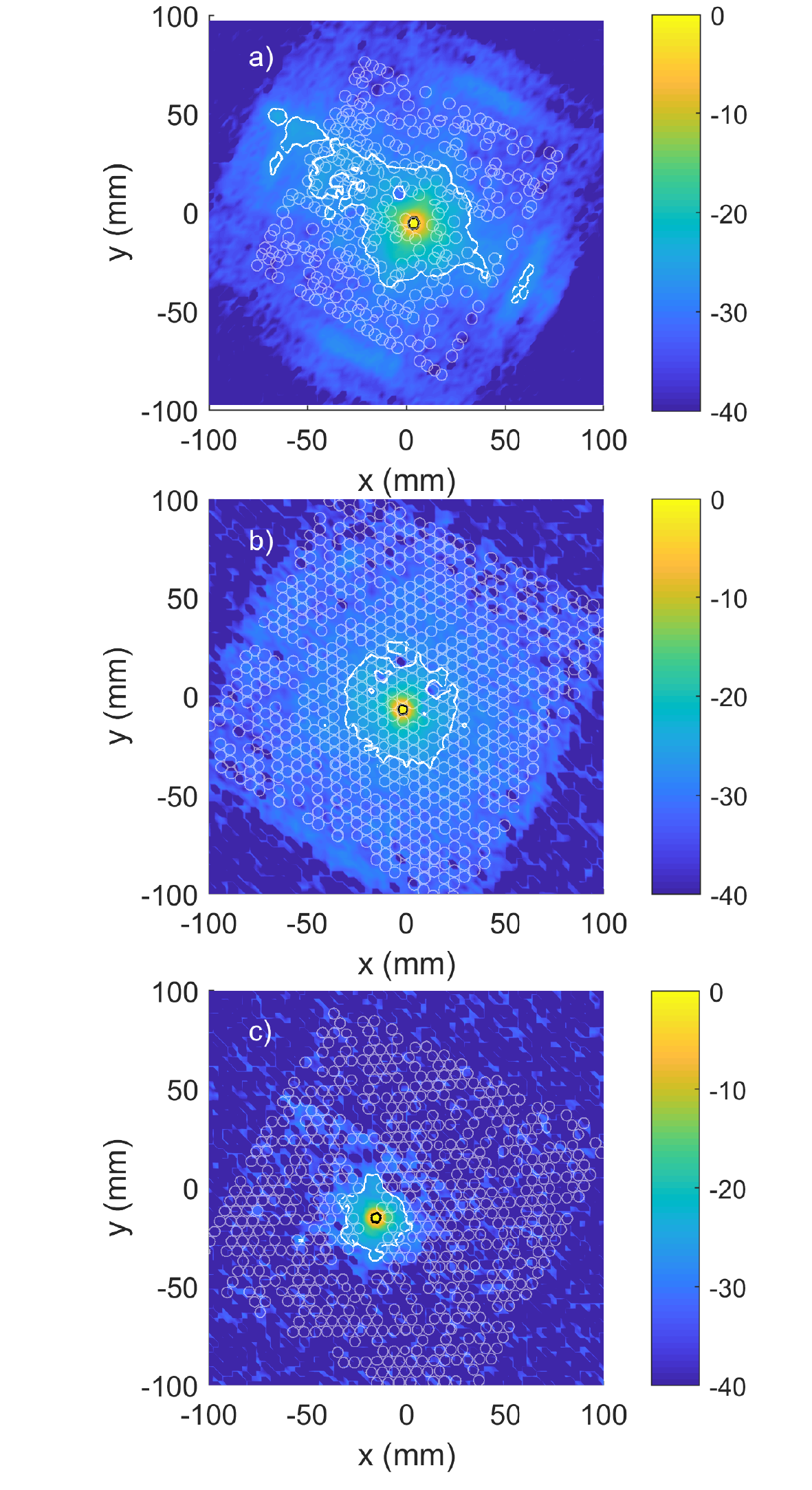}
\caption{Position dependent response in dB of one pixel to a point source placed in a reimaged focal plane with a magnification of 3. The $-3$~dB and $-27$~dB contours are shown. The circles show the fitted 3~dB beams of all found pixels, shown to show the extent of the array. Three arrays are shown: a) a bare LEKID array; b) a lens-antenna array without on chip stray light absorbing mesh; c) with absorbing mesh. Note the large area response at the $\sim-30$~dB level without the mesh disappears on the array with the on chip absorbing mesh.}
\label{fig:hot_beams}
\end{figure} 

The arrays were measured individually in a submillimeter wave camera cryostat, with the arrays mounted on a thermal isolation suspension connected to the 240~mK stage of a three stage He$^{3}$/He$^{3}$/He$^{4}$ sorption cooler. The two additional cold stages of this three stage cooler are used to thermally buffer the co-axial readout lines and thermal-mechanical suspension holding the detector assembly. The camera optics create an image of the detector array at a warm focal plane outside the cryostat using a seven mirror system with a system magnification of 3. The optical design is based on two back-to-back optical relays, each consisting two off-axis parabolic mirrors forming a Gaussian beam telescope~\cite{Goldsmith}. One of the optical relays is placed at 4~K and the other outside the cryostat. The optical design is based on aberration compensation~\cite{murphy:IJRMW87}, canceling the aberrations and cross-polarization of the optics near the optical axis. To improve performance over the entire (large) field of view the mirror shape and angles are optimized, giving a low distortion, diffraction limited performance with a Strehl ratio of greater than 0.97 across the entire field view at 350~GHz and even at 850~GHz. Three fold mirrors are used to minimize the total size of the optics system and give a horizontal beam with a usable warm reimaged focal plane. This rotates the focal plane, which is not corrected for in the presented data. An angular limiting aperture ``the pupil'' limits the beam to a focal length to beam diameter (f-number or f\#) of f\#=2 and is placed between the 4~K active mirrors where all the different pixel beams overlap. The optical band is selected with a bandpass filter at 350~GHz with a $\sim$~40~GHz bandwidth, while a selection of low pass and IR filters limit the out of band stray radiation~\cite{yates:IEEETTT18short}.

The arrays are read out using an in-house developed multiplexed readout system~\cite{rantwijk:IEEE16}, which allows 2~GHz of readout bandwidth to be readout. Separate RF cards up and down convert the readout around a central local oscillator (LO) frequency: one card allows between 5~GHz and 7~GHz  to be measured; a second card allows 2\ldots4~GHz to be readout. The lens-antenna MKIDs are designed to have resonant frequencies in range 4\ldots8~GHz with a frequency spacing that also scales with frequency. Since the central 50~MHz of the readout is not usable it takes 4 different LO tunings to measure the entire array. The LEKID array is designed to be readout out directly with the 2\ldots4~GHz RF card.

To measure the position-dependent response of the arrays we use a unpolarized hot source placed in the reimaged focal plane and scanned using a xy scanner. The hot source consists of a globar element placed in the focus of an enclosing elliptical mirror. The elliptical mirror produces an image of the source at its second focus, where a $\varnothing2$~mm beam-defining aperture is placed. The output of the source is modulated at 80~Hz between 300~K and upto 1000~K by means of a rotating mirror to eliminate MKID 1/f noise and system thermal drifts. The hot source has been previously characterized to have a wide response with an effective spot size smaller than the beamsizes to be measured. To maintain a constant background on the MKIDs and to eliminate reflections, a blackened sheet significantly larger than the reimaged chip size is mounted around the source aperture. The response of the MKID as a function of the source position is measured using a step-and-integrate strategy. The typical step size of the source is chosen to be close to spatial Nyquist sampling (2.5~mm) to enable an efficient sampling of the entire field of view. For each xy point, typically one second of data is recorded. The hot source temperature is adjusted so the maximum signal during the measurement matches the MKID instantaneous dynamic range. This is  taken in MKID phase readout~\cite{day03,gao:APL07} as $\sim$1~rad. with respect to the MKID resonance circle in the complex plane. In post processing, the data is calibrated to an effective frequency shift via the MKID phase signal using the strategy outlined in~\cite{bisigello:JLTP16}. This linearizes the signal with respect to the optical power and removes responsivity changes due to drifts in the optical loading. This results in reproducible beams for different source powers, MKID readout tone frequencies and powers. The position-dependent response is determined by applying a flat-top windowed FFT to each second of MKID frequency response and taking the FFT amplitude at the chopper frequency. Since the chopper is not locked it has a slight frequency drift. To correct for this, we modulate a few off-KID readout tones using analogue electronics with the exact chopper frequency. This gives the exact chopper frequency in the measured data, enabling drift correction in post-processing. This was implemented only for the array with absorber. For the arrays without absorber we use the sum of all MKIDs for this purpose: the pedestal response ensures that a signal is present independent of the source position. 

\begin{figure*}
\centering
\includegraphics*[width=17cm]{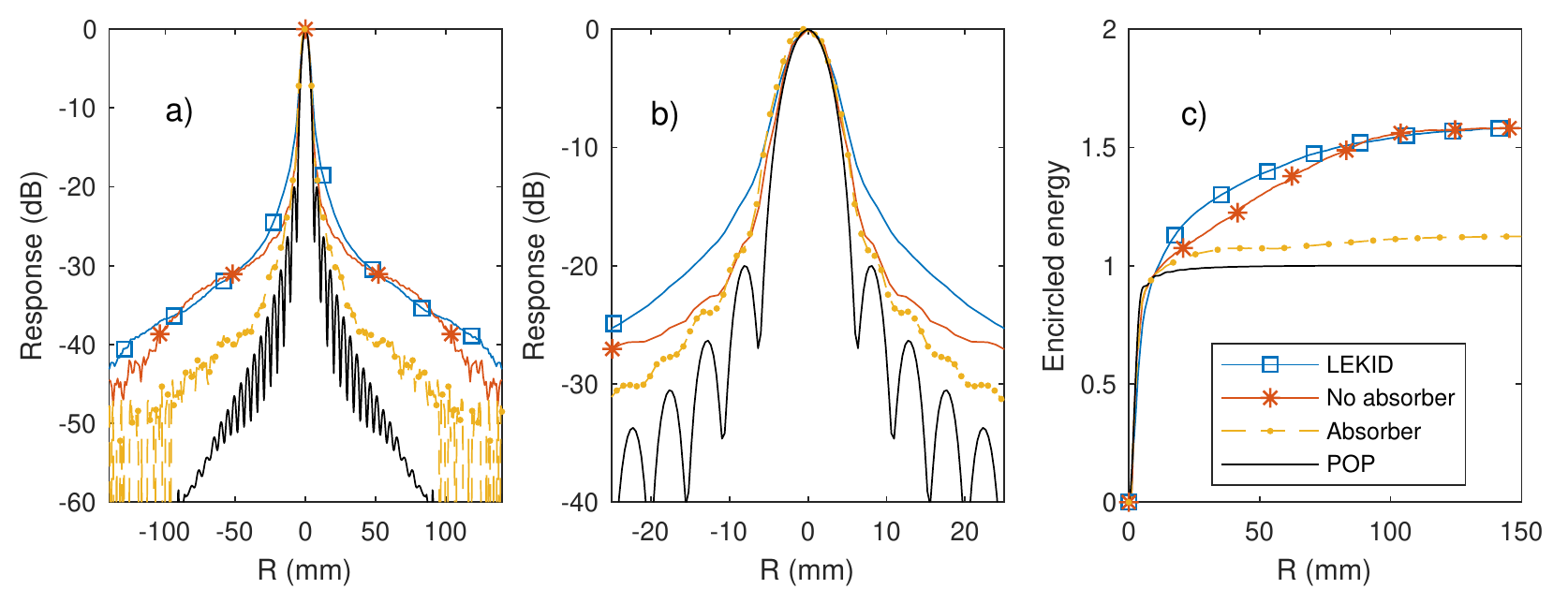}
\caption{ a) Radial mean of beam pattern. Shown is the median of $\sim20$ pixels from the array center. b) Zoom on the radial mean. c) Encircled energy, the integral of the beam pattern to a given radius centered on each pixel. The encircled energy is the median value of $\sim$ 20 central pixels, normalized to the total power of the POP simulation and corrected for slightly different beam radii (see text for details).}\label{fig:radial_mean}
\end{figure*}

The position dependent responses is shown: in Fig.~\ref{fig:hot_beams}(a) for the LEKID array; in Fig.~\ref{fig:hot_beams}(b) for the lens-antenna array without absorbing mesh and in Fig.~\ref{fig:hot_beams}(c) for the lens-antenna array with absorbing mesh. In all figures we give the corrected response $P_{c}$, given by $P_{c}=\sqrt{P^{2}-P_n^2}$, where $ P_n^2$ is the mean value of the signal with the hot source outside of the field of view of the cryostat optics. The square is needed because the KID noise, readout noise and photon noise contributions all add in $P^{2}$, with $P$ the measured signal. 

We observe a localized peak response, the main beam,  at the pixel position. However, for the LEKID and lens-antenna array without an on chip absorber we also observe a low-level of response over the entire chip area, which we will refer to as the "pedestal" response in the remainder of the text. The pedestal response consists of power coupled to the chip at a position spatially far away from the measured pixel: it is detected at the pixel under test due to scattering of radiation inside the detector chip. Normalizing the system beam pattern to its maximum response, the pedestal response is seen at a level of $\sim-30$~dB. In this particular case the total integrated stray power in the pedestal at $-30$~dB is similar to the power in the main beam. The pedestal level is source polarization dependent: in a separate measurement a polarized source reduces the pedestal level by $\sim3$~dB. We observe a significant reduction ($\sim10$~dB) in the pedestal response for the array with mesh absorber. To make this even clearer we show in Figs.~\ref{fig:radial_mean}(a)~\&~(b) the radial mean of the beam pattern. The radial mean averages the beam pattern over a circle centered on the pixel position, so improving the signal to noise of the beam pattern. The solid (blue) line, representing the data without a mesh absorber, is $\sim10$~dB above the (red) dashed line, representing the data with absorber, for radii in excess of 20~mm. For comparison, to indicate the expected level due to diffraction we also show in Figs.~\ref{fig:radial_mean} a physical optics (POP) simulation of the entire setup using a gaussian beam similar to lens-antenna beam pattern. Residual difference between simulation and measured patterns are attributed to the source beam pattern, defocus and optics quality~\cite{yates:IEEETTT18short}.

To illustrate the difference between arrays we show in Fig.~\ref{fig:radial_mean}(c) the encircled energy, which is the integral of the beam pattern over a circle centered on the pixel position. Note, the encircled energy is normalized to the simulation (POP) and corrected for small variations in the FWHM between the curves. The FWHM and hence its integral vary between measurements on the order of $\sim 10~\%$ due to the finite source size for hot source measurements, defocus, slight optical misalignments, and slight difference between measurements and simulation. Noting that the main beam response dominates up to a radius of 10~mm, the power inside this radius is used to normalize the encircled energy to the POP beam pattern. The encircled energy shows that without absorber there is almost 1.6~$\times$ the response of the array with an absorber, and that this extra power is distributed away from the pixel center. 

\section{Conclusions}
In this paper we have shown that large, monolithic arrays of lens antenna MKIDs and LEKIDs can respond to radiation, on a $-30$~dB level, over the entire chip area. This "pedestal" response is associated with stray radiation scattered inside the dielectric of the array substrate, it is commonly referred to as a surface wave. The integrated response of the pedestal approaches the main beam response. Such a response destroys the imaging properties of the array, particularly for extended sources. Measurements on a LEKID array shows similar pedestal performance to a lens-antenna array. The effect is associated with radiation falling on the array not directly absorbed in the detector, but scattered in the substrate. As such, it should scale with the inverse square of total array optical efficiency: due to a higher direct pixel response which in addition gives less stray power to be absorbed. This problem occurs even when the individual pixel has a high optical efficiency such as when using single moded detectors in a spatial undersampled configuration so reducing the array efficiency~\cite{Griffin:AO02}, as in the presented lens-antenna arrays. 

We have shown~\cite{yates:IEEETTT18short} for lens-antenna designs that the surface wave can be suppressed effectively by including a matched absorbing layer in between the detector chip and the lens array. The absorbing layer reduces the surface wave by at least 10~dB. For LEKID designs, a solution is less clear, but higher efficiency design than presented should suppress it but not remove it. Note that no array is 100\% efficient, for example due to finite filling fraction, while ensuring high efficiency over a wide band and high frequency has its own issues~\cite{NIKA2:AA18short}. However, a lens-coupled LEKID design could implement a similar on chip stray light absorber as presented. Horn coupled devices on a transparent substrate are also sensitive to stray coupling via the substrate and so also integrate on chip stray light absorbing layer~\cite{McCarrick:JLTP2016short} .

\acknowledgments % equivalent to \section*{ACKNOWLEDGMENTS}       
This work was in part supported by ERC starting grant ERC-2009-StG Grant 240602 TFPA. The contribution form N. Llombart is supported by the  European Research Council Starting Grant LAA-THz-CC (639749). The contribution of J.J.A. Baselmans is also supported by the ERC consolidator grant COG 648135 MOSAIC. This work was part of a collaborative project, SPACEKIDs, funded via grant 313320 provided by the European Commission under Theme SPA.2012.2.2-01 of Framework Programme 7.

% References

\bibliography{Steve,general_refs}

\begin{thebibliography}{10}

\bibitem{NIKA2:AA18short}
{Adam, R.} et~al., ``The {NIKA2} large-field-of-view millimetre continuum
  camera for the 30 m iram telescope,'' {\em A\&A}~{\bf 609},  A115 (2018).

\bibitem{amkid}
G\"uesten, R., Heyminck, S., et~al., ``{A-MKID}.''

\bibitem{Brien:LTD18}
{Brien}, T.~L.~R., {Castillo-Dominguez}, E., {Chase}, S., and {Doyle}, S.~M.,
  ``{A Continuous 100-mK Helium-Light Cooling System for MUSCAT on the LMT},''
  {\em ArXiv e-prints}  (Jan. 2018).

\bibitem{CMBS4reviewshort}
{Abazajian}, K.~N. et~al., ``{CMB-S4 Science Book, First Edition},'' {\em ArXiv
  e-prints}  (Oct. 2016).

\bibitem{Doyle:JLTP08}
Doyle, S., Mauskopf, P., Naylon, J., Porch, A., and Duncombe, C., ``Lumped
  element kinetic inductance detectors,'' {\em J. Low Temp. Phys.}~{\bf 151},
  530--536 (Apr 2008).

\bibitem{yates:IEEETTT18short}
Yates, S. J.~C. et~al., ``Surface wave control for large arrays of microwave
  kinetic inductance detectors,'' {\em IEEE Trans. THz Sci. Technol.}~{\bf 7},
  789--799 (Nov 2017).

\bibitem{day03}
Day, P., LeDuc, H., Mazin, B., Vayonakis, A., and Zmuidzinas, J., ``A broadband
  superconducting detector suitable for use in large arrays,'' {\em
  Nature}~{\bf 425},  817--821 (OCT 23 2003).

\bibitem{omid:IEEEMTT12}
Noroozian, O., Day, P.~K., Eom, B.~H., Leduc, H.~G., and Zmuidzinas, J.,
  ``Crosstalk reduction for superconducting microwave resonator arrays,'' {\em
  IEEE Trans. on Microw. Theory and Techn.}~{\bf 60},  1235--1243 (May 2012).

\bibitem{Janssen:APL13}
Janssen, R. M.~J., Baselmans, J. J.~A., Endo, A., Ferrari, L., Yates, S. J.~C.,
  Baryshev, A.~M., and Klapwijk, T.~M., ``High optical efficiency and photon
  noise limited sensitivity of microwave kinetic inductance detectors using
  phase readout,'' {\em Appl. Phys. Lett.}~{\bf 103}(20),  203503 (2013).

\bibitem{ferrari:IEEETTT18}
Ferrari, L., Yurduseven, O., Llombart, N., Yates, S. J.~C., Bueno, J.,
  Murugesan, V., Thoen, D.~J., Endo, A., Baryshev, A.~M., and Baselmans, J.
  J.~A., ``Antenna coupled mkid performance verification at 850 ghz for large
  format astrophysics arrays,'' {\em IEEE Trans. THz Sci. Technol.}~{\bf 8},
  127--139 (Jan 2018).

\bibitem{Jonas:IEEETMTT92}
Zmuidzinas, J. and LeDuc, H.~G., ``Quasi-optical slot antenna {SIS} mixers,''
  {\em IEEE Trans. on Microw. Theory Techn.}~{\bf 40},  1797--1804 (Sep 1992).

\bibitem{filipovic:IEEETMTT93}
Filipovic, D.~F., Gearhart, S.~S., and Rebeiz, G.~M., ``Double-slot antennas on
  extended hemispherical and elliptical silicon dielectric lenses,'' {\em IEEE
  Trans. on Microw. Theory Techn.}~{\bf 41},  1738--1749 (Oct 1993).

\bibitem{Gao2008b}
Gao, J., Daal, M., Martinis, J.~M., Vayonakis, A., Zmuidzinas, J., Sadoulet,
  B., Mazin, B.~A., Day, P.~K., and Leduc, H.~G., ``A semiempirical model for
  two-level system noise in superconducting microresonators,'' {\em Appl. Phys.
  Lett.}~{\bf 92}(21),  212504 (2008).

\bibitem{Ozan2016}
Yurduseven, O., {\em Wideband integrated lens antennas for terahertz deep space
  investigation}, PhD thesis, Dept. Microelectronics, Delft University of
  Technology , Delft, The Netherlands (2016).

\bibitem{Yates:JLTP14}
Yates, S. J.~C., Baselmans, J. J.~A., Baryshev, A.~M., Doyle, S., Endo, A.,
  Ferrari, L., Hochg{\"u}rtel, S., and Klein, B., ``{Clean Beam Patterns with
  Low Crosstalk Using 850 GHz Microwave Kinetic Inductance Detectors},'' {\em
  J. Low Temp. Phys.}~{\bf 176}(5),  761--766 (2014).

\bibitem{Adane:JLTP16}
Adane, A., Boucher, C., Coiffard, G., Leclercq, S., Schuster, K.~F., Goupy, J.,
  Calvo, M., Hoarau, C., and Monfardini, A., ``{Crosstalk in a KID Array Caused
  by the Thickness Variation of Superconducting Metal},'' {\em J. Low Temp.
  Phys.}~{\bf 184},  137--141 (Jul 2016).

\bibitem{Bisigello:SPIE16}
Bisigello, L., Yates, S. J.~C., Ferrari, L., Baselmans, J. J.~A., and Baryshev,
  A., ``Measurements and analysis of optical crosstalk in a microwave kinetic
  inductance detector array,'' {\em Proc. SPIE}~{\bf 9914},  99143L (2016).

\bibitem{baselmans:AA17short}
{Baselmans, J. J. A.} et~al., ``A kilo-pixel imaging system for future space
  based far-infrared observatories using microwave kinetic inductance
  detectors,'' {\em A\&A}~{\bf 601},  A89 (2017).

\bibitem{Thoen2016}
Thoen, D.~J., Bos, B. G.~C., Haalebos, E. A.~F., Klapwijk, T.~M., Baselmans, J.
  J.~A., and Endo, A., ``Superconducting {NbTiN} thin films with highly uniform
  properties over a ${varnothing}$ 100 mm wafer,'' {\em IEEE Trans. on Appl.
  Supercond.}~{\bf 27},  1--5 (June 2017).

\bibitem{schrey1970}
Schrey, F., Mathis, R., and Payne, R., ``Structure and properties of rf
  sputtered, superconducting tantalum films,'' {\em Thin Solid Films}~{\bf
  5}(1),  29--40 (1970).

\bibitem{mohazzab2000}
Mohazzab, M., Mulders, N., Nash, A., and Larson, M., ``Tantalum thin-film
  superconducting transition edge thermometers,'' {\em J. Low Temp. Phys.}~{\bf
  121}(5-6),  821--824 (2000).

\bibitem{Griffin:AO02}
Griffin, M.~J., Bock, J.~J., and Gear, W.~K., ``Relative performance of filled
  and feedhorn-coupled focal-plane architectures,'' {\em Appl. Opt.}~{\bf 41},
  6543--6554 (Nov 2002).

\bibitem{Goldsmith}
Goldsmith, P.~F.,  [{\em Gaussian Beam Quasioptical Propagation and
  Applications}{\nolinebreak\hspace{0.1em}]}, IEEE Press (1998).

\bibitem{murphy:IJRMW87}
Murphy, J.~A., ``Distortion of a simple gaussian beam on reflection from
  off-axis ellipsoidal mirrors,'' {\em Int. J. Infrared and Milli. Waves}~{\bf
  8},  1165--1187 (Sep 1987).

\bibitem{rantwijk:IEEE16}
van Rantwijk, J., Grim, M., van Loon, D., Yates, S., Baryshev, A., and
  Baselmans, J., ``Multiplexed readout for 1000-pixel arrays of microwave
  kinetic inductance detectors,'' {\em IEEE Trans. Microw. Theory Techn.}~{\bf
  64},  1876--1883 (June 2016).

\bibitem{gao:APL07}
Gao, J., Zmuidzinas, J., Mazin, B.~A., LeDuc, H.~G., and Day, P.~K., ``Noise
  properties of superconducting coplanar waveguide microwave resonators,'' {\em
  Appl. Phys. Lett.}~{\bf 90}(10) (2007).

\bibitem{bisigello:JLTP16}
Bisigello, L., Yates, S. J.~C., Murugesan, V., Baselmans, J. J.~A., and
  Baryshev, A.~M., ``Calibration scheme for large kinetic inductance detector
  arrays based on readout frequency response,'' {\em J. Low Temp. Phys.}~{\bf
  184}(1),  161--166 (2016).

\bibitem{McCarrick:JLTP2016short}
McCarrick, H. et~al., ``A titanium nitride absorber for controlling optical
  crosstalk in horn-coupled aluminum lekid arrays for millimeter wavelengths,''
  {\em J. of Low Temp. Phys.}~{\bf 184},  154--160 (Jul 2016).

\end{thebibliography}
\bibliographystyle{spiebib} % makes bibtex use spiebib.bst

\end{document}